\documentclass[11pt,aps,showpacs,floatfix,onecolumn,tightenlines]{revtex4}
\usepackage{epsfig}
\usepackage{psfig}
\usepackage{graphicx}
\usepackage{graphics}
\usepackage{xspace}
\usepackage{amssymb}
\usepackage{amsmath}
\usepackage{latexsym}
\usepackage{natbib}
\usepackage{mathrsfs}
\newcommand{\inieq}{\begin{eqnarray}}            
\newcommand{\fineq}{\end{eqnarray}}            
\newcommand{\diff}{{\rm\,d}}                    

\def\p{\mbox{\boldmath $p$}}

\def\q{\mbox{\boldmath $q$}}

\def\mcg{\mbox{$\mathcal{G}$}}
\def\mcv{\mbox{$\mathcal{V}$}}
\begin{document}
\title{Relativistic approach to neutrino-nucleus quasielastic scattering} 

\author{Andrea Meucci}
\author{Carlotta Giusti} 
\author{Franco Davide Pacati }
\affiliation{Dipartimento di Fisica Nucleare e Teorica, 
Universit\`{a} di Pavia, \\ and
Istituto Nazionale di Fisica Nucleare, 
Sezione di Pavia, I-27100 Pavia, Italy}


\begin{abstract}  A relativistic Green's function and a distorted-wave 
impulse-approximation approach to charged- and neutral-current neutrino-nucleus 
quasielastic scattering are developed.
Results for the neutrino (antineutrino) reactions on $^{16}$O and
$^{12}$C target nuclei
are presented and discussed.\end{abstract} 
 \maketitle

\section{Introduction}
\label{sec.int}

We are here interested in $\nu(\bar\nu)$-nucleus 
scattering reactions in the quasielastic (QE) region, where the neutrino
interacts with one single nucleon and the mechanisms like RPA or the 
excitations of the target nucleus become less important. 
In Sec. \ref{sec2} a Green's function approach to describe inclusive 
charged-current (cc) neutrino scattering is developed \cite{cc}. 
In such a process, $\nu(\bar\nu)$'s interact with nuclei
via the exchange of weak-vector bosons and charged leptons are produced in the
final state.
In our model 
the conservation of flux is preserved and final state interactions are treated 
consistently with an exclusive reaction. 
In Sec. \ref{sec3} a relativistic distorted-wave impulse-approximation
(RDWIA) calculation of semi-inclusive neutral-current (nc) 
$\nu$- and $\bar\nu$-nucleus reactions 
is presented \cite{nc}. The sensitivity to
the strange quark content of the nucleon weak current is discussed.

\section{The relativistic Green's function method}
\label{sec2}

The cross section of an inclusive reaction where an incident neutrino or
antineutrino interacts with a nucleus and only the outgoing lepton is detected 
is given by the contraction 
between the lepton tensor \cite{cc} and the hadron
tensor, whose components are given by bilinear products of the transition 
matrix elements of the nuclear weak charged-current operator $J^{\mu}$. 
The current operator is assumed to be adequately described as 
the sum of single-nucleon currents, corresponding to the weak charged current  
\begin{eqnarray}
  j^{\mu} = \big[ F_1^{{\mathrm V}}
   \gamma ^{\mu} + 
        i\frac {\kappa}{2M} F_2^{{\mathrm V}}
	\sigma^{\mu\nu}q_{\nu}
	 - G_{{\mathrm A}}
	 \gamma ^{\mu}\gamma ^{5} +
	     F_{{\mathrm P}}
	     q^{\mu}\gamma ^{5} \big] \tau^{\pm}\ ,
	     \label{eq.cc}
\end{eqnarray}
where $\tau^{\pm}$ are the isospin operators, $\kappa$ is the anomalous part of 
the magnetic moment, $q^{\mu} = (\omega , \q)$, with $Q^2 = |\q|^2 - \omega^2$, 
is the four-momentum transfer, and
$\sigma^{\mu\nu}=\left(i/2\right)\left[\gamma^{\mu},\gamma^{\nu}\right]$.
$F_1^{{\mathrm V}}$ and $F_2^{{\mathrm V}}$ are the isovector 
Dirac and Pauli 
nucleon form factors. 
$G_{{\mathrm A}}$ and $F_{{\mathrm P}}$ are the axial and 
induced pseudoscalar form factors, which are usually parametrized as
\begin{eqnarray}
G_{{\mathrm A}}(Q^2) =  g_{{{\mathrm A}}} 
        \left(1+Q^2/M^2_{{\mathrm A}}\right)^{-2} \ , \quad
F_{{\mathrm P}}(Q^2)= 2MG_{{\mathrm A}} \left(m^2_{\pi} 
+ Q^2\right)^{-1} \ , \label{eq.formf}
\end{eqnarray}
where $g_{{\mathrm A}} = 1.267$, $m_{\pi}$ is the pion mass, 
and $M_{{\mathrm A}} \simeq (1.026 \pm 0.021)$ GeV is the axial mass. 
Performing the contraction between the lepton and hadron 
tensors, the inclusive cross section for the QE $\nu$($\bar\nu$)-nucleus 
scattering can be written as \cite{Walecka}
\begin{eqnarray}
\frac{\diff \sigma} {\diff \varepsilon \diff \Omega} = 
 \frac{k\varepsilon  G^2} {4 \pi^2}
  \cos^2\vartheta_{{\mathrm c}} 
  \Big[ v_0 R_{00} + v_{zz} R_{zz} - v_{0z} R_{0z} + v_T R_T 
  \pm v_{xy} R_{xy} \Big]    ,
\label{eq.cs}
\end{eqnarray} 
where the coefficients $v$ are obtained from the lepton tensor components. 
All nuclear structure information is contained in the response
functions $R$, which are defined in terms of the hadron tensor components, i.e.,
\inieq
 W^{\mu\nu}(\omega,q)  
=\langle \Psi_0 \mid J^{\nu\dagger}(\q) \delta(E_{\textrm {f}}-H) 
J^{\mu}(\q) \mid \Psi_0 \rangle \ . 
\label{eq.hadrontensor}
\fineq
Introducing the Green's operator 
related to the nuclear Hamiltonian $H$, we have 
\inieq
 \omega^{\mu\mu} =  W^{\mu\mu}(\omega,q)  
= -\frac{1}{\pi} \textrm{Im} \langle \Psi_0 \mid J^{\mu\dagger}(\q) 
G(E_{\textrm {f}}) J^{\mu}(\q) \mid \Psi_0 \rangle \ , 
\label{eq.ht1}
\fineq
for $\mu = 0,x,y,z$, and simliar expressions for $\omega^{0z}$ and
$\omega^{xy}$ \cite{cc}.
 
It was shown in Refs. \cite{cc,ee} that the nuclear response in 
Eq. (\ref {eq.hadrontensor}) can be written in terms of 
the
single particle Green's function, $\mcg(E)$, whose self-energy is the 
Feshbach's optical potential. 
A biorthogonal 
expansion of the full particle-hole Green's operator in terms of the 
eigenfunctions of the non-Hermitian optical potential
$\mcv$, and of its Hermitian conjugate $\mathcal{V}^{\dagger}$, is performed
\inieq
\left[ {\mathcal{E}} - T - {\mathcal{V}}^{\dagger} (E) \right] \mid
{\chi}_{\mathcal{E}}^{(-)}(E)\rangle = 0\ , \ \ \ \ \  
\left[ \mathcal{E} - T - {\mathcal{V}}(E) \right] \mid \tilde
{\chi}_{\mathcal{E}}^{(-)}(E)\rangle = 0\ .
 \label{eq.op}
\fineq
Note that $E$ and ${\mathcal{E}}$ are not necessarily the same.
The spectral representation is
\inieq
\mcg(E) = \int_M^{\infty} \diff \mathcal{E}\mid\tilde
{\chi}_{\mathcal{E}}^{(-)}(E)\rangle 
 \frac{1}{E-\mathcal{E}+i\eta} \langle\chi_{\mathcal{E}}^{(-)}(E)\mid 
\ . \label{eq.sperep}
\fineq
The hadron tensor components can be written in an expanded form  in terms of the
single-particle wave function, $\mid \varphi_n \rangle$, of the initial state,
corresponding to the energy $\varepsilon_n$ and whose spectral strength is
$\lambda_n$ as
\inieq
\omega^{\mu\nu}(\omega , q) = -\frac{1}{\pi} \sum_n  \textrm{Im} \bigg[
 \int_M^{\infty} \diff \mathcal{E} \frac{1}{E_{{\mathrm
{f}}}-\varepsilon_n-\mathcal{E}+i\eta}  
  T_n^{\mu\nu}(\mathcal{E} ,E_{{\mathrm{f}}}-\varepsilon_n) \bigg]
\ , \label{eq.pracw}
\fineq
where
\inieq
T_n^{\mu\mu}(\mathcal{E} ,E) &=& \lambda_n\langle \varphi_n
\mid j^{\mu\dagger}(\q) \sqrt{1-\mcv'(E)}
\mid\tilde{\chi}_{\mathcal{E}}^{(-)}(E)\rangle \nonumber \\
&\times&  \langle\chi_{\mathcal{E}}^{(-)}(E)\mid  \sqrt{1-\mcv'(E)} j^{\mu}
(\q)\mid \varphi_n \rangle  \ , \label{eq.tprac}
\fineq
for $\mu = 0,x,y,z$, and simliar expressions for $\omega^{0z}$ and
$\omega^{xy}$ \cite{cc}. The factor 
 $\sqrt{1-\mcv'(E)}$ accounts for
interference effects between different channels and allows the replacement of
the mean field $\mcv$ by the phenomenological optical potential 
$\mcv_{ \mathrm L}$.   
After calculating the limit for $\eta \rightarrow +0$, 
Eq. \ref{eq.pracw} reads
\inieq
\omega^{\mu\nu}(\omega , q) = \sum_n \Bigg[ \textrm{Re} T_n^{\mu\nu}
(E_{\mathrm{f}}-\varepsilon_n, E_{ \mathrm{f}}-\varepsilon_n)  \nonumber
\\
- \frac{1}{\pi} \mathcal{P}  \int_M^{\infty} \diff \mathcal{E} 
\frac{1}{E_{\mathrm{f}}-\varepsilon_n-\mathcal{E}} 
\textrm{Im} T_n^{\mu\nu}
(\mathcal{E},E_{\mathrm{f}}-\varepsilon_n) \Bigg] \ , \label{eq.finale}
\fineq
where $\mathcal{P}$ denotes the principal value of the integral. 

Disregarding the square root correction, due to interference effects,
The second matrix element in Eq. \ref{eq.tprac}, with the inclusion of 
$\sqrt{\lambda_n}$ is the transition amplitude for the single-nucleon 
knockout 
from a nucleus in the state $\mid \Psi_0\rangle$ leaving the residual nucleus 
in the state $\mid n \rangle$. The attenuation of its strength,
mathematically due to the imaginary part of the optical potential, is related to 
the
flux lost towards the channels different from $n$. In the inclusive response
this attenuation must be compensated by a corresponding gain, due to the flux
lost, towards the channel $n$, by the other final states asymptotically
originated by the channels different from $n$. 
This compensation is
performed by the first matrix element in the right hand side of 
Eq. \ref{eq.tprac}, where the imaginary part of the potential has the effect of 
increasing the strength. Similar considerations can be made, on the purely 
mathematical ground, for the integral of Eq. \ref{eq.finale}, where the 
amplitudes involved in $T_n^{\mu\nu}$ have no evident physical meaning when 
${\mathcal{E}}\neq E_{\rm{f}}-\varepsilon_n$. 

In an usual shell-model calculation the cross section is obtained from the sum, 
over all the single-particle shell-model states, of the squared absolute value
of the transition matrix elements.
Therefore, in such a calculation the negative imaginary part of the
optical potential produces a loss of flux that is inconsistent with the 
inclusive process.
In the Green's function approach, the flux is conserved, as the
components of the hadron tensor are obtained in terms of the product of the two
matrix elements in Eq. \ref{eq.tprac}: 
the loss of flux, produced by the
negative imaginary part of the optical potential in $\chi$, is compensated by 
the gain of flux, produced in the first matrix element by the positive 
imaginary part of the Hermitian conjugate optical potential in $\tilde \chi$. 
The cross sections and the response functions of the inclusive QE
$\nu$($\bar\nu$)-nucleus scattering are calculated from the 
single-particle expression of the hadron tensor in  Eq. \ref{eq.finale}. 
After the replacement of the mean field $\mcv(E)$ by the empirical optical 
model potential $\mcv_{\mathrm {L}}(E)$, the matrix elements of the nuclear 
current operator in Eq. \ref{eq.tprac} are of the same kind as those giving 
the transition 
amplitudes  of the electron induced nucleon knockout reaction \cite{book} and
the same RDWIA treatment can be used \cite{meucci1,meucci2}.
 
As a study case, we have considered the $^{16}$O target nucleus and two 
different values of the incident neutrino energy 
$E_\nu$ = 500 and 1000 MeV. 
In order to show up the effect of the optical potential on the inclusive 
reaction, the results obtained in the present approach are compared with those 
given by different approximations. 
In the simplest one the optical potential is neglected and the plane 
wave approximation is assumed for the final state wave functions ${\chi}^{(-)}$
and  $\tilde{\chi}^{(-)}$. In this plane wave impulse approximation (PWIA) FSI
between the outgoing nucleon and the residual nucleus are completely neglected. 
In another approach the imaginary part of the optical potential is neglected 
and only the real part is included. 
This approximation conserves the flux, but it is inconsistent with the 
exclusive 
process, where a complex optical potential must be used. Moreover, the use of 
a real optical potential is unsatisfactory from a theoretical point of view, 
since the optical potential has to be complex owing to the presence of open 
channels. 
\begin{figure}[t]
\includegraphics[height=12cm, width=12cm]{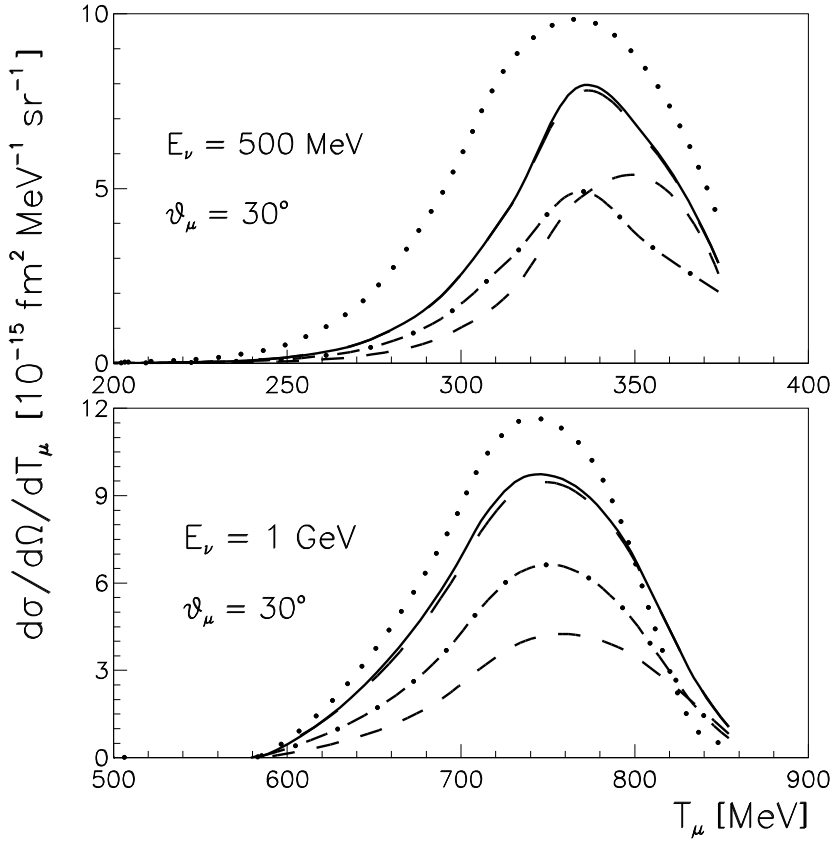}
\vskip -.2cm
\caption {The differential cross sections of the $^{16}$O$(\nu_{\mu},\mu^-)$ 
reaction for $E_\nu$ = 500 and 1000 MeV at $\vartheta_\mu$ = 30 degrees.  
Solid lines represent the result of the Green's function approach,
dotted lines give PWIA, long-dashed lines show the result with
a real optical potential, and dot-dashed lines the contribution of 
the integrated exclusive reactions with one-nucleon emission.
Short dashed lines give the cross sections of the 
$^{16}$O$(\bar\nu_{\mu},\mu^+)$ reaction calculated with the Green's 
function approach. }
\label{fig1}
\end{figure}
In Fig. \ref{fig1}, the differential cross sections of the 
$^{16}$O$(\nu_{\mu},\mu^-)$ reaction for $\vartheta_\mu$ = 30 degrees are 
displayed as a function of the muon kinetic energy $T_\mu$. 
The behavior of the calculated cross sections is similar for the different 
energies. The effect of the optical potential increases with $T_\mu$ and 
decreases increasing $E_\nu$. The result of the PWIA is about 20--30\% higher 
at the peak than the one of the Green's function approach. 
The sum of the exclusive one-nucleon emission cross sections is always much 
smaller than the complete result. The difference indicates the relevance of
inelastic channels and is due to the loss of flux produced by the absorptive 
imaginary part of the optical potential. In contrast, the cross sections 
calculated  with only the real part of the optical potential are practically 
the same as the ones obtained with the Green's function approach. Although the 
use of a complex optical potential is conceptually important from a 
theoretical point of view, the negligible differences given by the two results 
mean that the conservation of flux, that is fulfilled in both calculations, is 
the most important condition in the present situation. In contrast, significant 
differences are obtained with a real optical potential in the inclusive 
electron scattering~\cite{ee}. The cross sections for 
the $^{16}$O$(\bar\nu_{\mu},\mu^+)$ reaction are also shown for a comparison. 
They are always much smaller than the corresponding cross sections with an
incident neutrino.

\section{The neutral-current semi-inclusive scattering}
\label{sec3}

The differential cross section for the neutral-current 
$\nu$($\bar\nu$)-nucleus quasielastic scattering is obtained from the 
contraction between the lepton and hadron tensors, as in Ref. \cite{Walecka}. 
After performing an integration over the solid angle of the 
final nucleon, we have
\begin{eqnarray}
{  \frac{\diff \sigma}  
{\diff \varepsilon \diff \Omega \diff {\mathrm {T_N}}}} &=& 
 \frac{G^2\varepsilon^2} {2 \pi^2}  \cos^2 \frac {\vartheta}{2} 
  \Big [ v_{0} R_{00} + v_{zz} R_{zz} - v_{0z} R_{0z} + v_T R_T 
  \pm v_{xy} R_{xy} \Big] \frac {|\p_{\mathrm N}| E _{\mathrm N}} {(2 \pi)^3}
 \ ,
\label{eq.ncs}
\end{eqnarray} 
where $\vartheta$ is the lepton scattering angle and $E _{\mathrm N}
(\p_{\mathrm N})$ the energy (momentum) of the outgoing nucleon. 
The coefficients $v$ and the
responses $R$ are obtained
from the lepton and hadron tensor components, respectively \cite{nc}.

The transition matrix elements are calculated in the first
order perturbation theory and in the impulse approximation. Thus, the 
transition amplitude is assumed to be adequately 
described as the sum of terms similar to those appearing in the electron
scattering case \cite{book,meucci1}.
The RDWIA treatment is the same 
as in Refs.~\cite{meucci1,meucci2}. 
The single-particle current operator related to the weak neutral current is  
\begin{eqnarray}
  j^{\mu} =  F_1^{\mathrm V} \gamma ^{\mu} + 
      i\frac {\kappa}{2M} F_2^{ \mathrm V}\sigma^{\mu\nu}q_{\nu}	  
	     -G_{ \mathrm A}\gamma ^{\mu}\gamma ^{5} +
	     F_{ \mathrm P}q^{\mu}\gamma ^{5}\ ,
	     \label{eq.nc}
\end{eqnarray}
The vector form factors $F_i^{\mathrm V}$ can be expressed in terms of the 
corresponding electromagnetic form factors for protons $(F_i^{\mathrm p})$ and 
neutrons $(F_i^{\mathrm n})$, plus a possible isoscalar strange-quark 
contribution $(F_i^{\mathrm s})$, i.e.,
\begin{eqnarray}
F_i^{ \mathrm V;\ p(n)} = \pm \left\{F_i^p - F_i^n\right\}/2 - 
2\sin^2{\theta_{\mathrm W}}F_i^{p(n)} - F_i^s/2\ ,
\end{eqnarray}
where $+(-)$ stands for proton (neutron) knockout and $\theta_{\mathrm W}$ is 
the Weinberg angle $(\sin^2{\theta_{\mathrm W}} 
\simeq 0.2313)$. 
The strange vector form factors are taken as \cite{gar93}
\begin{eqnarray}
F_2^{\mathrm s}(Q^2)  =  \frac {F_2^{\mathrm s}(0)} {(1+\tau) 
(1+Q^2/M_{\mathrm V}^2)^2} , \  
F_1^{\mathrm s}(Q^2) =  \frac {F_1^{\mathrm s} Q^2}{(1+\tau) 
(1+Q^2/M_{\mathrm V}^2)^2}  ,
\label{eq.sform}
\end{eqnarray}
where $\tau = Q^2/(4M_p^2)$, $F_2^{\mathrm s}(0) = \mu_{\mathrm s}$, 
$F_1^{\mathrm s} = -\langle r^2_{\mathrm s}\rangle /6$, and $M_{\mathrm V}$ =
0.843 GeV. The quantity $\mu_{\mathrm s}$ is the strange magnetic moment and 
$\langle r^2_{\mathrm s}\rangle$ the squared {\lq\lq strange radius\rq\rq} 
of the nucleon. 
The axial form factor is expressed as 
\begin{eqnarray}
G_{{\mathrm A}}(Q^2)  =  \frac {  \pm g_{{\mathrm A}}^{\phantom{ }} - 
g_{{{\mathrm A}}}^s}{2\left(1+Q^2/M^2_{{\mathrm A}}\right)^{2}}    \ , 
\end{eqnarray}
where  
$g^{\mathrm s}_{\mathrm A}$ describes possible strange-quark contributions.

\begin{figure}[ht]
\includegraphics[height=12cm, width=12cm]{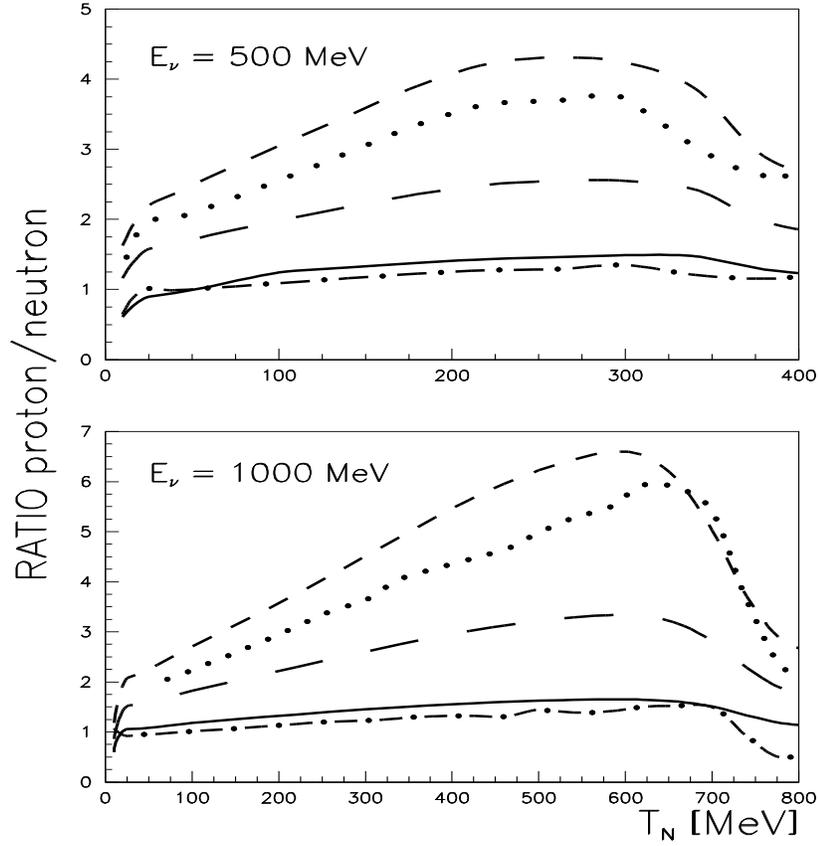}
\vskip -0.2cm
\caption {Ratio of proton to neutron total cross sections of the $\nu (\bar \nu)$
quasielastic scattering on $^{12}$C as a function of the
incident neutrino (antineutrino) energy. 
Solid and dashed lines are the results in RDWIA with 
$g^{\mathrm s}_{\mathrm A} = 0$ and $g^{\mathrm s}_{\mathrm A} = -0.19$. 
Dot-dashed and dotted lines are the same results but for an incident
antineutrino. Long-dashed line corresponds to neutrino scattering with 
$g^{\mathrm s}_{\mathrm A} = -0.10$.
}
\label{fig2}
\end{figure}

\begin{figure}[ht]
\includegraphics[height=12cm, width=12cm]{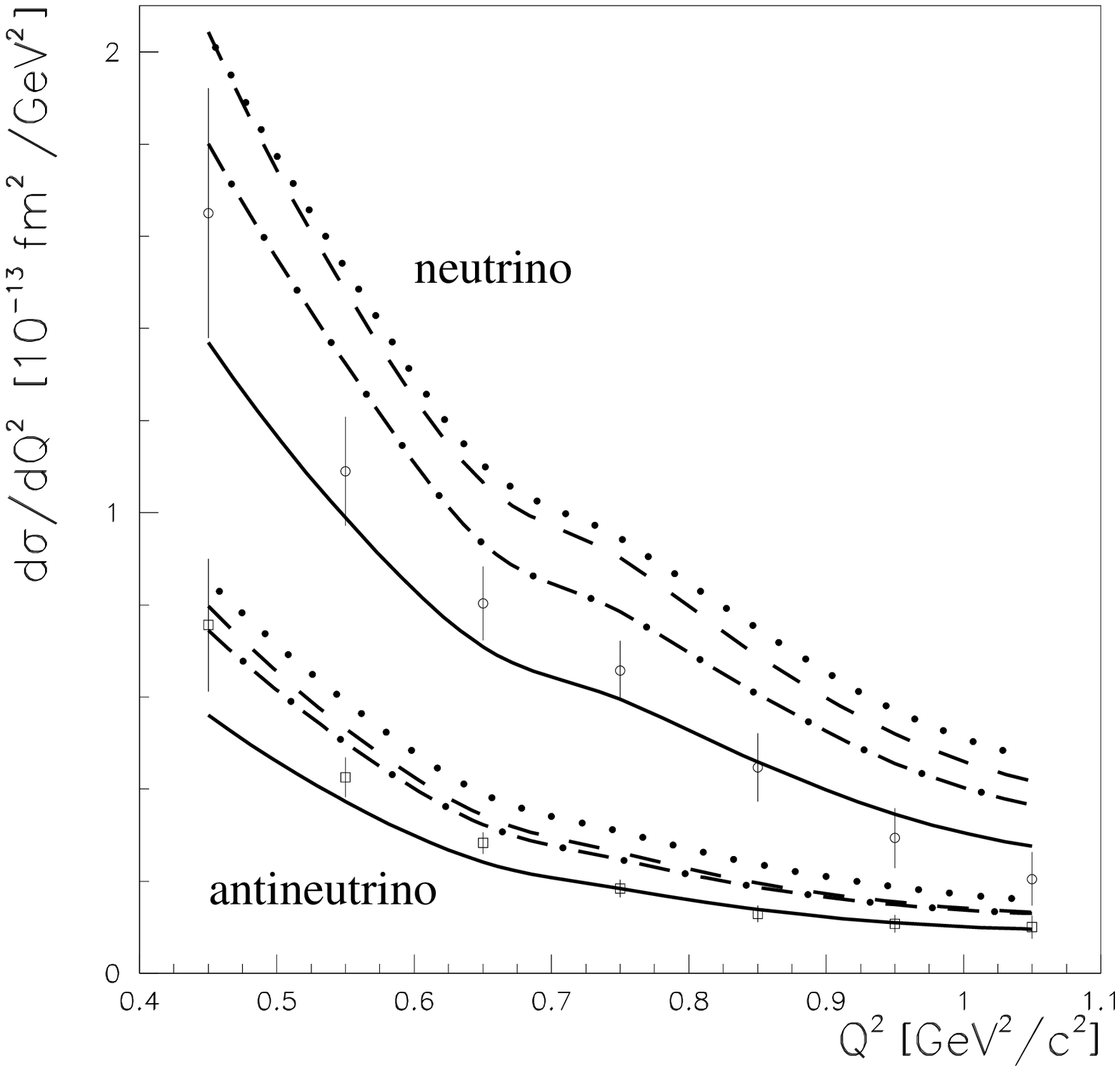}
\vskip -0.2cm
\caption {Differential cross sections of the $\nu (\bar \nu)$
quasielastic scattering, flux-averaged over BNL 
spectrum \cite{bnl}, as a function of the momentum transfer squared. The four 
upper curves
are for incident neutrino and the four lower ones for incident antineutrino.
Solid lines are the results with no strangeness 
contribution, 
dashed lines with $g^{\mathrm s}_{\mathrm A} = -0.19$, dot-dashed lines
with $g^{\mathrm s}_{\mathrm A} = -0.19$ and $ F_2^{\mathrm s}(0) = -0.40$,
dotted lines with $g^{\mathrm s}_{\mathrm A} = -0.19$, 
$ F_2^{\mathrm s}(0) = -0.40$ and
$F_1^{\mathrm s} =- \langle r^2_{\mathrm s}\rangle /6 = 0.53$ GeV$^{-2}$.
Experimental data from Ref. \cite{bnl}. The errors correspond to statistical
and $Q^2$-dependent systematic errors added in quadrature and do not include
$Q^2$-independent systematic errors.
}
\label{fig3}
\end{figure}

In order to separate the effects of the strange-quark contribution and of FSI 
on the cross sections, it was suggested in Refs. \cite{gar92} to 
measure the ratio of proton to neutron yields, as this ratio is
expected to be less sensitive to distortion effects than the cross sections
themselves. Moreover, from the experimental point of view the ratio is less 
sensitive to the uncertainties in the determination of the incident neutrino 
flux. In Fig. \ref{fig2} the ratio for an incident neutrino is
displayed as a function of the outgoing nucleon kinetic energy both in RDWIA 
and RPWIA. 
The ratio is very sensitive to $g^{\mathrm s}_{\mathrm A}$ and exhibits a 
maximum at $T_N \simeq 0.6 \ E_\nu$, which, however, corresponds to values 
where the cross sections are small. The effect is sensibly reduced for 
$g^{\mathrm s}_{\mathrm A} = -0.10 $ with respect to 
$g^{\mathrm s}_{\mathrm A} = -0.19 $.  
An enhancement of the ratio of $\simeq 15$\% is produced by FSI. 
This result is due both to Coulomb distortion and to the different coupling
of the optical potential with proton and neutron currents. 
It means that the argument of looking
for the strange-quark content in this ratio is strengthened by distortion, but
the possibility to fix the exact value of the contribution is affected by the
uncertainties due to FSI.

Finally, in Fig. \ref{fig3}  we compare our results with the data of the 
BNL 734 experiment \cite{bnl}. Experimental results were presented in the form of
a flux-averaged differential cross section per momentum transfer squared $Q^2$.
Our results are shown with $g^{\mathrm s}_{\mathrm A} = -0.19$
and without the strange-quark contribution. 
We give also the effect of including the
strange vector form factors, $F_1^{\mathrm s}$ and $F_2^{\mathrm s}$, with 
$F_1^{\mathrm s} =-\langle r^2_{\mathrm s}\rangle /6  = 0.53$ GeV$^{-2}$, 
$ F_2^{\mathrm s}(0) = -0.40$ \cite{gar}, and the $Q^2$ dependence given in 
Eq. \ref{eq.sform}. 
The strange-quark contribution produces an enhancement of the cross sections, 
which makes them slightly higher than the experimental data. The strange weak
magnetic contribution decreases the cross section, while the axial and weak
electric components give an enhancement.


\begin{thebibliography}{}




\bibitem{cc}
A. Meucci, C. Giusti, and F.D. Pacati, 
{ Nucl. Phys. A} {\bf 739}, 277 (2004).

\bibitem{nc}
A. Meucci, C. Giusti, and F.D. Pacati, 
{ Nucl. Phys. A} {\bf 744}, 307 (2004).

\bibitem{Walecka}
J.D. Walecka, in {\sl Muon Physics}, Vol. II, edited by V.H. Hughes and 
C.S. Wu (Academic Press, New York, 1975), p. 113. 

\bibitem{ee}
A. Meucci, F. Capuzzi, C. Giusti, and F.D. Pacati,
 { Phys. Rev. C} {\bf 67}, 054601 (2003).

\bibitem{book}
S. Boffi, {\it et al.},
{\it Electromagnetic Response of Atomic Nuclei}, Oxford Studies in Nuclear
Physics, Vol. 20 (Clarendon, Oxford, 1996);
S. Boffi, {\it et al.}, Phys. Rep. {\bf 226}, 1 (1993).

\bibitem{meucci1}
A. Meucci, C. Giusti, and F.D. Pacati, 
 {Phys. Rev. C} {\bf 64}, 014604 (2001).

\bibitem{meucci2}
A. Meucci, C. Giusti, and F.D. Pacati, 
{ Phys. Rev. C}  {\bf 64}, 064615 (2001).

\bibitem{gar93}
G.T. Garvey, {\it et al.},
 { Phys. Rev. C}  {\bf 48}, 1919 (1993).

\bibitem{gar92}
G.T. Garvey, {\it et al.},
{ Phys. Lett. B} {\bf 289}, (1992) 249; 
C.J. Horowitz, {\it et al.}, { Phys. Rev. C}  {\bf  48}, 3078 (1993); 
W.M. Alberico, {\it et al.},
{ Nucl. Phys. A} {\bf 623}, 471 (1997); 
W.M. Alberico, {\it et al.}, { Phys. Lett. B} {\bf  438}, 9 (1998).

\bibitem{bnl}
L.A. Ahrens, {\sl et al.},
{ Phys. Rev. D}  {\bf 35}, 785 (1987).

\bibitem{gar} 
G.T. Garvey, {\it et al.},
 { Phys. Rev. C} {\bf 48}, 761 (1993).




\end{thebibliography}
\end{document}